\documentclass{article}
\usepackage{spconf,amsmath,amssymb,graphicx}

\title{RVQ Position Aware Speculative Decoding for On Device
Text to Speech}

\name{Berkin Durmus, Eduardo Pacheco, Zach Nagengast, Atila Orhon}
\address{Argmax, Inc., USA}

\renewcommand{\paragraph}[1]{\par\noindent\textbf{#1}}
\newcommand{\rev}[1]{#1}
\newcommand{\revb}[1]{#1}

\usepackage{comment}
\usepackage{bookmark}
\usepackage{caption}
\usepackage{float}
\usepackage{multirow}
\usepackage{tikz}
\usepackage{pgfplots}
\pgfplotsset{compat=1.18}
\usetikzlibrary{positioning,arrows.meta,shapes.geometric,calc,fit,backgrounds,decorations.pathreplacing,decorations.pathmorphing}

\begin{document}
\ninept
\setlength{\parskip}{0.15\baselineskip plus 1pt minus 2pt}
\setlength{\textfloatsep}{7pt plus 2pt minus 3pt}
\setlength{\floatsep}{7pt plus 2pt minus 2pt}
\setlength{\intextsep}{7pt plus 2pt minus 2pt}
\setlength{\abovecaptionskip}{3pt}
\makeatletter
\renewcommand\section{\@startsection{section}{1}{\z@}%
  {-2.4ex plus -1ex minus -.2ex}{1.5ex plus .2ex}{\large\bf}}
\renewcommand\subsection{\@startsection{subsection}{2}{\z@}%
  {-2.2ex plus -1ex minus -.2ex}{1.1ex plus .2ex}{\normalsize\bf}}
\makeatother
\let\origthebibliography\thebibliography
\renewcommand{\thebibliography}[1]{\origthebibliography{#1}%
  \setlength{\itemsep}{0.5pt plus 0.5pt}\setlength{\parskip}{0pt}}

\maketitle

\begin{abstract}
Autoregressive decoding (AR) with Transformer models is
memory bandwidth bound at single stream inference, the typical
deployment regime for on device text to speech (TTS). Real time
streaming with Qwen3-TTS~\cite{qwen3tts} requires $\geq 200$
sequential model calls per second, dominated by the inner loop
\texttt{MultiCodeDecoder} that emits the $15$
residual vector quantization (RVQ) codes per $80$\,ms audio
frame. We propose \emph{RVQ position aware speculative decoding}
for the \texttt{MultiCodeDecoder}, attaining $2.47$ accepted
tokens per model call at \revb{$5{\times}10^{-4}\%$} added parameters and
$10$--$20\%$ \revb{per round} speculation/verification overhead,
\revb{reducing real time synthesis from $200$ to ${\approx}88$
sequential model calls per second}.
\revb{The scheme is distributionally lossless under the deployed
top-$k$ sampling~\cite{leviathan2023fast,chen2023accelerating},
and WER parity with the original system is consistent with this
guarantee.} We deliver
$2$--$2.2\times$ speedup for RVQ token generation with
Qwen3-TTS-0.6B \revb{on recent} iPhone and Apple Silicon
Mac devices.
\end{abstract}


\section{Introduction}
\label{sec:intro}

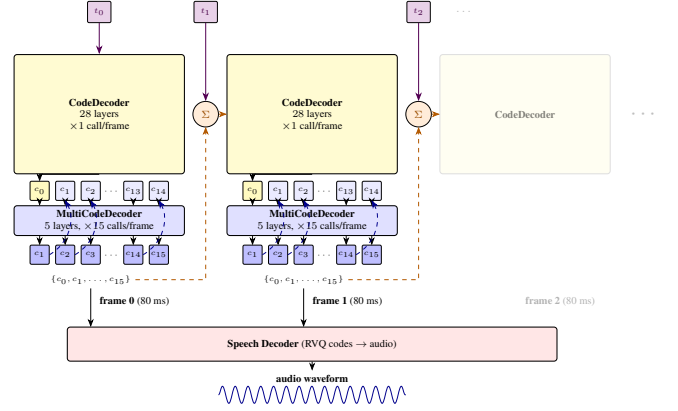
\begin{figure}[t]
\centering
\resizebox{\columnwidth}{!}{%
\begin{tikzpicture}[
    every node/.style={font=\scriptsize},
    cdblk/.style={draw,rounded corners=2pt,minimum width=40mm,minimum height=28mm,
                  inner sep=2pt,fill=yellow!22,align=center,line width=0.7pt},
    mcdblk/.style={draw,rounded corners=2pt,minimum width=40mm,minimum height=7mm,
                   inner sep=1pt,fill=blue!12,align=center,line width=0.7pt},
    cseed/.style={draw,rounded corners=1pt,minimum width=4.5mm,minimum height=4.5mm,
                  inner sep=0pt,fill=yellow!30,line width=0.4pt,font=\tiny},
    cin/.style={draw,rounded corners=1pt,minimum width=4.5mm,minimum height=4.5mm,
                inner sep=0pt,fill=blue!8,line width=0.4pt,font=\tiny},
    cout/.style={draw,rounded corners=1pt,minimum width=4.5mm,minimum height=4.5mm,
                 inner sep=0pt,fill=blue!25,line width=0.4pt,font=\tiny},
    ttok/.style={draw,rounded corners=1pt,minimum width=5.5mm,minimum height=5mm,
                 inner sep=0pt,fill=violet!18,line width=0.4pt,font=\tiny,
                 text=violet!55!black},
    sumblk/.style={draw,circle,minimum size=6mm,inner sep=0pt,
                   fill=orange!15,line width=0.5pt,font=\scriptsize,
                   text=orange!60!black},
    spkrblk/.style={draw,rounded corners=2pt,minimum height=8mm,inner sep=2pt,
                    fill=red!10,align=center,line width=0.7pt},
    arr/.style={->,>=Stealth,thin},
    fbarr/.style={->,>=Stealth,thin,dashed,orange!70!black},
    innerloop/.style={->,>=Stealth,thin,dashed,blue!55!black},
    inputarr/.style={->,>=Stealth,semithick,violet!65!black}
  ]
  \def\dx{5.0}

  \node[ttok] (tt0) at (0,       4.4) {$t_0$};
  \node[ttok] (tt1) at (\dx/2,   4.4) {$t_1$};
  \node[ttok] (tt2) at (\dx*1.5, 4.4) {$t_2$};
  \node[font=\scriptsize,gray!60] at ([xshift=8mm]tt2.east) {$\cdots$};

  \node[cdblk] (cd0) at (0, 2)
    {\textbf{CodeDecoder}\\$28$ layers\\$\times 1$ call/frame};

  \node[cseed] (f0i0)  at (-1.4, 0.2) {$c_0$};
  \node[cin]   (f0i1)  at (-0.8, 0.2) {$c_1$};
  \node[cin]   (f0i2)  at (-0.2, 0.2) {$c_2$};
  \node[font=\tiny] at (0.3, 0.2) {$\cdots$};
  \node[cin]   (f0i13) at (0.8, 0.2) {$c_{13}$};
  \node[cin]   (f0i14) at (1.4, 0.2) {$c_{14}$};

  \node[mcdblk] (mcd0) at (0, -0.5)
    {\textbf{MultiCodeDecoder}\\$5$ layers, $\times 15$ calls/frame};

  \node[cout] (f0o1)  at (-1.4, -1.3) {$c_1$};
  \node[cout] (f0o2)  at (-0.8, -1.3) {$c_2$};
  \node[cout] (f0o3)  at (-0.2, -1.3) {$c_3$};
  \node[font=\tiny] at (0.3, -1.3) {$\cdots$};
  \node[cout] (f0o14) at (0.8, -1.3) {$c_{14}$};
  \node[cout] (f0o15) at (1.4, -1.3) {$c_{15}$};

  \draw[arr] (cd0.south) -| (f0i0.north);
  \foreach \n in {f0i0,f0i1,f0i2,f0i13,f0i14}
    \draw[arr] (\n.south) -- (\n.south |- mcd0.north);
  \foreach \n in {f0o1,f0o2,f0o3,f0o14,f0o15}
    \draw[arr] (\n.north |- mcd0.south) -- (\n.north);
  \draw[innerloop] (f0o1.east)  to[out=30,in=-60] (f0i1.south);
  \draw[innerloop] (f0o2.east)  to[out=30,in=-60] (f0i2.south);
  \draw[innerloop] (f0o14.east) to[out=30,in=-60] (f0i14.south);

  \node[font=\tiny,below=1.5mm of f0o3,align=center] (rvq0)
    {$\{c_0,c_1,\dots,c_{15}\}$};

  \draw[inputarr] (tt0.south) -- (cd0.north);

  \node[sumblk] (sum0) at (\dx/2, 2) {$\Sigma$};
  \draw[inputarr] (tt1.south) -- (sum0.north);
  \draw[fbarr] (rvq0.east) -| (sum0.south);

  \node[cdblk] (cd1) at (\dx, 2)
    {\textbf{CodeDecoder}\\$28$ layers\\$\times 1$ call/frame};

  \node[cseed] (f1i0)  at (\dx-1.4, 0.2) {$c_0$};
  \node[cin]   (f1i1)  at (\dx-0.8, 0.2) {$c_1$};
  \node[cin]   (f1i2)  at (\dx-0.2, 0.2) {$c_2$};
  \node[font=\tiny] at (\dx+0.3, 0.2) {$\cdots$};
  \node[cin]   (f1i13) at (\dx+0.8, 0.2) {$c_{13}$};
  \node[cin]   (f1i14) at (\dx+1.4, 0.2) {$c_{14}$};

  \node[mcdblk] (mcd1) at (\dx, -0.5)
    {\textbf{MultiCodeDecoder}\\$5$ layers, $\times 15$ calls/frame};

  \node[cout] (f1o1)  at (\dx-1.4, -1.3) {$c_1$};
  \node[cout] (f1o2)  at (\dx-0.8, -1.3) {$c_2$};
  \node[cout] (f1o3)  at (\dx-0.2, -1.3) {$c_3$};
  \node[font=\tiny] at (\dx+0.3, -1.3) {$\cdots$};
  \node[cout] (f1o14) at (\dx+0.8, -1.3) {$c_{14}$};
  \node[cout] (f1o15) at (\dx+1.4, -1.3) {$c_{15}$};

  \draw[arr] (cd1.south) -| (f1i0.north);
  \foreach \n in {f1i0,f1i1,f1i2,f1i13,f1i14}
    \draw[arr] (\n.south) -- (\n.south |- mcd1.north);
  \foreach \n in {f1o1,f1o2,f1o3,f1o14,f1o15}
    \draw[arr] (\n.north |- mcd1.south) -- (\n.north);
  \draw[innerloop] (f1o1.east)  to[out=30,in=-60] (f1i1.south);
  \draw[innerloop] (f1o2.east)  to[out=30,in=-60] (f1i2.south);
  \draw[innerloop] (f1o14.east) to[out=30,in=-60] (f1i14.south);

  \node[font=\tiny,below=1.5mm of f1o3,align=center] (rvq1)
    {$\{c_0,c_1,\dots,c_{15}\}$};

  \node[sumblk] (sum1) at (\dx*1.5, 2) {$\Sigma$};
  \draw[inputarr] (tt2.south) -- (sum1.north);
  \draw[fbarr] (rvq1.east) -| (sum1.south);

  \node[cdblk,opacity=0.5,draw=gray!50,fill=yellow!12] (cd2) at (\dx*2, 2)
    {\textbf{CodeDecoder}};
  \node[font=\Large,gray!50] at ([xshift=8mm]cd2.east) {$\cdots$};

  \draw[fbarr] (sum0.east) |- (cd1.west);
  \draw[fbarr] (sum1.east) |- (cd2.west);

  \node[font=\scriptsize,anchor=west] (t0lab) at (-0.1, -2.4)
    {\textbf{frame 0} (80 ms)};
  \node[font=\scriptsize,anchor=west] (t1lab) at (\dx-0.1, -2.4)
    {\textbf{frame 1} (80 ms)};
  \node[font=\scriptsize,gray!60,anchor=west] (t2lab) at (\dx*2-0.1, -2.4)
    {\textbf{frame 2} (80 ms)};

  \node[spkrblk,minimum width=115mm] (spkr) at (\dx, -3.4)
    {\textbf{Speech Decoder} (RVQ codes $\to$ audio)};
  \draw[arr] (rvq0.south) -- (rvq0.south |- spkr.north);
  \draw[arr] (rvq1.south) -- (rvq1.south |- spkr.north);

  \node[font=\scriptsize,below=2mm of spkr,align=center] (wave)
    {\textbf{audio waveform}};
  \draw[arr] (spkr.south) -- (wave.north);
  \draw[blue!55!black,line width=0.55pt,decorate,
        decoration={snake,amplitude=2mm,segment length=3mm}]
    ([xshift=-22mm,yshift=-1.8mm]wave.south)
    -- ([xshift=22mm,yshift=-1.8mm]wave.south);
\end{tikzpicture}}
\caption{\textbf{Qwen3-TTS autoregressive audio generation.} A
forced text-token sequence $t_0, t_1, t_2, \dots$ (violet) is
fed into the model: each $t_i$ is an input text token consumed 
by frame $i$. $t_0$ conditions the first \emph{CodeDecoder}; for
$i\!>\!0$, $t_i$ is summed with the $16$ RVQ codes of frame $i\!-\!1$
inside $\Sigma$ to condition the next \emph{CodeDecoder}.
At each frame the \emph{CodeDecoder} is called once
and emits the seed code $c_0$ (yellow); the \emph{MultiCodeDecoder}
is then called autoregressively $15$ times: at inner step $k$ it
consumes $c_{k-1}$ and emits $c_k$, with each output fed back as the
next-step input, completing the $16$-code RVQ frame embedding
$\{c_0,\dots,c_{15}\}$, which is then summed with $t_i$ in $\Sigma$.
\emph{Speech Decoder} consumes the resulting RVQ code stream and
renders the audio waveform. Real-time synthesis requires $12.5\,\text{frames/s}\times16
\,\text{codes/frame}=200$ tokens/s.}
\label{fig:rvqloop}
\end{figure}

Speech waveforms carry complex timbre and prosodic information in
addition to the underlying linguistic content. As a result,
state of the art text to speech (TTS) systems, e.g.\
Qwen3-TTS~\cite{qwen3tts}, CosyVoice~3~\cite{cosyvoice}, and
CaT-TTS~\cite{cattts}, are designed to emit many more tokens per
second of audio than per character of input text. As shown in
Figure~\ref{fig:rvqloop}, Qwen3-TTS uses two nested loops over
residual vector quantization (RVQ)
codes\rev{~\cite{soundstream2021,encodec2022}}: the outer loop runs a
text conditioned \texttt{CodeDecoder} once per $80$\,ms audio
frame to emit the coarse code $c_0$, and the inner loop runs the
\texttt{MultiCodeDecoder} $15$ times \emph{autoregressively} to
emit the $15$ residual codes $c_1,\dots,c_{15}$ that complete the
embedding of that frame. In total, ${\sim}1$\,s of audio is
represented with ${\sim}200$ tokens: $13$ outer loop
\texttt{CodeDecoder} calls and $187$ inner loop
\texttt{MultiCodeDecoder} calls.

Real time streaming generation therefore requires a minimum of
$200$ tokens per second of inference throughput, and, because
every token is produced by a sequential model call, this
translates into $\geq 200$ sequential model calls per second. The
additional throughput required by the audio token to waveform
decoder is negligible relative to that of token generation. On
single stream on device inference, the inner
\texttt{MultiCodeDecoder} loop dominates per frame latency,
because kernel launch overhead and KV cache management amortize
poorly across so many sequential calls.

Speculative decoding~\cite{leviathan2023fast,chen2023accelerating}
is a natural fit: a cheap drafter proposes $k$ candidate tokens
that the target verifies in one parallel model call; when on
average $\bar{k}$ candidates are accepted, the inner loop emits
$\bar{k}{+}1$ tokens per model call instead of $1$. We propose
RVQ position aware speculative decoding, which extends
Medusa~\cite{cai2024medusa} with a position aware hidden state
offset and a sparse tree attention mask so that drafter and
verifier fuse into a single compiled \texttt{MultiCodeDecoder}
model call (Figure~\ref{fig:offset}). The method adds ${<}0.001\%$
parameters and $10$--$20\%$ \revb{per round} speculation/verification
overhead, and \revb{is distributionally lossless under the
deployed top-$k$ sampling~\cite{fan2018}, inheriting the
guarantee of speculative
sampling~\cite{leviathan2023fast,chen2023accelerating}; WER
parity with the original system is consistent with this
guarantee.} Putting it
together, \revb{real time generation still emits $200$ tokens per
second but now requires only ${\approx}88$ sequential model calls
per second ($12.5$ outer plus ${\approx}76$ inner at $2.47$
tokens/step), and we demonstrate}
$2$--$2.2\times$ practical speedup for RVQ token generation
\revb{on recent} iPhone and Apple Silicon Mac devices.
Concretely, our method:


\begin{itemize}\itemsep -1pt
\item extends the Medusa framework\revb{~\cite{cai2024medusa}} with an RVQ position aware
      offset that fuses drafter and verifier into a single compiled
      \texttt{MultiCodeDecoder} model call
      (\S\ref{sec:system}, Figure~\ref{fig:offset}),
\item \revb{achieves WER parity with the autoregressive baseline
      across six languages, with no audible differences in manual
      review or production use, in line with the lossless guarantee
      of speculative
      sampling~\cite{leviathan2023fast,chen2023accelerating}
      (\S\ref{ssec:wer}),}
\item \rev{delivers $2.47$ accepted tokens per
      \texttt{MultiCodeDecoder} model call (\S\ref{ssec:tps}) and
      $2$--$2.2\times$ measured RVQ token generation speedup
      \revb{on recent devices such as
      iPhone 17 and MacBook M3} (\S\ref{ssec:device}), unlocking
      reliable real time TTS on older hardware such as
      iPhone 12.}
\end{itemize}

\rev{\paragraph{Related work.}
Codec language models~\cite{valle2023} cast TTS as autoregressive
token generation over RVQ codes. In this setting,
VADUSA~\cite{vadusa2024} and Nguyen et
al.~\cite{mtpspeech2025} apply Medusa style~\cite{cai2024medusa}
multihead drafting to AR speech synthesis, but both accelerate the
temporal token loop, train new multimillion parameter draft heads,
verify with relaxed or heuristic acceptance rules that alter the
output distribution, and report GPU only results;
general purpose drafters such as EAGLE~\cite{eagle2024} and
ReDrafter~\cite{redrafter2024} likewise train new draft networks
on top of the backbone. On device, these designs carry material
overhead: a separate draft model or tens of millions of added
parameters increase asset size, memory footprint, and per call
compute, which mobile hardware can ill afford.
\revb{Principled coarse grained acceptance~\cite{pcg2026} instead
relaxes exact token matching to acoustic similarity groups,
trading token level exactness for acceptance; we preserve the
exact token level distribution.} Non speculative
alternatives such as delay pattern interleaving~\cite{musicgen2023} and
masked parallel generation~\cite{soundstorm} also reduce sequential
RVQ calls, but require retraining the backbone and change its output
distribution, which is undesirable for an already deployed model. We
instead target the RVQ depth inner loop and exploit the
\texttt{MultiCodeDecoder}'s native per position multihead structure,
which Medusa style methods must otherwise train from scratch:
drafting adds only \revb{$3{,}072$} trained offset parameters and no new
heads, \revb{verification preserves the deployed top-$k$ sampling
distribution}, and, extending our WhisperKit
inference stack~\cite{whisperkit}, drafter and verifier fuse into a
single static shape model call; to our knowledge this is the first
speculative decoding for TTS demonstrated on the Apple Neural
Engine.}

\section{System}
\label{sec:system}

\subsection{Medusa as the natural baseline}
Medusa~\cite{cai2024medusa} attaches $K$ extra Language Model
Heads (LMHs) to a target backbone so that one model call on
hidden state $h$ produces a draft of the next $K$ tokens. Unlike
conventional transformer decoders, the Qwen3-TTS
\texttt{MultiCodeDecoder} is multiheaded by construction: each
of the \revb{$15$ residual} codebook positions has its own dedicated
input embedding table and its own LM head, so the
previously committed code at position $i$ is looked up in table
$i$ and the hidden state at position $i$ is projected by
$\mathrm{LMH}_i$. This is a natural basis for Medusa style
drafting. A direct
application of Medusa would attach $K$ \emph{additional} drafter
heads, but the existing per position heads already cover the tokens
we want to speculate on. Reusing them is preferable: it adds no
parameters, requires no extra asset variants, and leaves the
output distribution unchanged.

\subsection{Achieving higher tokens per step}
\label{ssec:levers}

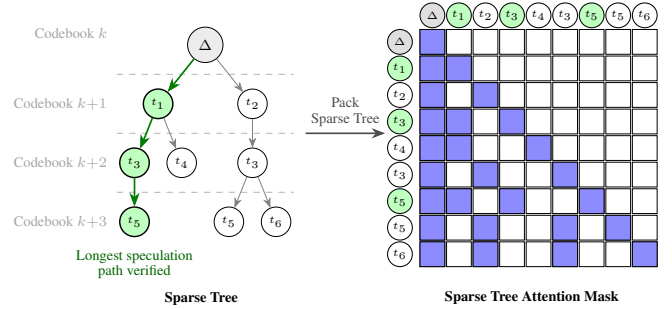
\begin{figure}[h]
\centering
\resizebox{\columnwidth}{!}{%
\begin{tikzpicture}[
    every node/.style={font=\footnotesize},
    n/.style={draw,circle,minimum size=5mm,inner sep=0pt,fill=white,
              font=\tiny},
    a/.style={draw,circle,minimum size=5mm,inner sep=0pt,
              fill=green!25,line width=0.7pt,font=\tiny},
    arr/.style={->,>=Stealth,thin,gray}
  ]
  \node[draw,circle,minimum size=6mm,fill=gray!15,inner sep=0pt] (r)
    at (0,0) {\scriptsize $\Delta$};
  \node[font=\scriptsize,gray!70,anchor=east] at (-1.55,0.2)
    {Codebook $k$};
  \draw[dashed,gray!60] (-1.5,-0.50) -- (1.5,-0.50);
  \node[font=\scriptsize,gray!70,anchor=east] at (-1.55,-1.0)
    {Codebook $k{+}1$};
  \draw[dashed,gray!60] (-1.5,-1.50) -- (1.5,-1.50);
  \node[font=\scriptsize,gray!70,anchor=east] at (-1.55,-2.0)
    {Codebook $k{+}2$};
  \draw[dashed,gray!60] (-1.5,-2.50) -- (1.5,-2.50);
  \node[font=\scriptsize,gray!70,anchor=east] at (-1.55,-3.0)
    {Codebook $k{+}3$};

  \node[a] (d11) at (-0.8,-1.0) {$t_1$};
  \node[n] (d12) at ( 0.8,-1.0) {$t_2$};
  \draw[arr] (r) -- (d11);
  \draw[arr] (r) -- (d12);

  \node[a] (e20) at (-1.2,-2.0) {$t_3$};
  \node[n] (e21) at (-0.4,-2.0) {$t_4$};
  \node[n] (e30) at ( 0.8,-2.0) {$t_3$};
  \draw[arr] (d11) -- (e20);
  \draw[arr] (d11) -- (e21);
  \draw[arr] (d12) -- (e30);

  \node[a] (f20) at (-1.2,-3.0) {$t_5$};
  \node[n] (f30) at ( 0.4,-3.0) {$t_5$};
  \node[n] (f31) at ( 1.2,-3.0) {$t_6$};
  \draw[arr] (e20) -- (f20);
  \draw[arr] (e30) -- (f30);
  \draw[arr] (e30) -- (f31);

  \draw[->,>=Stealth,thick,green!50!black] (r)   -- (d11);
  \draw[->,>=Stealth,thick,green!50!black] (d11) -- (e20);
  \draw[->,>=Stealth,thick,green!50!black] (e20) -- (f20);

  \node[align=center,below=1mm of f20,font=\scriptsize,
        green!40!black]
    {Longest speculation\\path verified};

  \node[font=\scriptsize,anchor=west] at (-0.8,{-0.15-8*0.45-0.30 -0.25})
    {\textbf{Sparse Tree}};

  \draw[->,>=Stealth,thick,black!70]
    (1.7,-1.5) -- (3.05,-1.5)
    node[midway,above,font=\scriptsize,align=center]
      {Pack\\Sparse Tree};

  \begin{scope}[shift={(3.85,0.2)}]
    \foreach \i/\lab/\fill in {%
      0/$\Delta$/gray!25,
      1/$t_1$/green!25,
      2/$t_2$/white,
      3/$t_3$/green!25,
      4/$t_4$/white,
      5/$t_3$/white,
      6/$t_5$/green!25,
      7/$t_5$/white,
      8/$t_6$/white%
    } {
      \node[draw,circle,minimum size=4.2mm,inner sep=0pt,fill=\fill,
            line width=0.5pt,font=\tiny]
        at ({\i*0.45},0.30) {\lab};
    }

    \foreach \i/\lab/\fill in {%
      0/$\Delta$/gray!25,
      1/$t_1$/green!25,
      2/$t_2$/white,
      3/$t_3$/green!25,
      4/$t_4$/white,
      5/$t_3$/white,
      6/$t_5$/green!25,
      7/$t_5$/white,
      8/$t_6$/white%
    } {
      \node[draw,circle,minimum size=4.2mm,inner sep=0pt,fill=\fill,
            line width=0.5pt,font=\tiny]
        at (-0.55,{-0.15-\i*0.45}) {\lab};
    }

    \foreach \r in {0,...,8} {
      \foreach \c in {0,...,8} {
        \node[draw,minimum size=4.2mm,inner sep=0pt,fill=white,
              line width=0.35pt]
          at ({\c*0.45},{-0.15-\r*0.45}) {};
      }
    }
    \foreach \r/\anc in {%
      0/{0},
      1/{0,1},
      2/{0,2},
      3/{0,1,3},
      4/{0,1,4},
      5/{0,2,5},
      6/{0,1,3,6},
      7/{0,2,5,7},
      8/{0,2,5,8}%
    } {
      \foreach \c in \anc {
        \node[draw,minimum size=4.2mm,inner sep=0pt,
              fill=blue!45,line width=0.35pt]
          at ({\c*0.45},{-0.15-\r*0.45}) {};
      }
    }

    \node[font=\scriptsize,anchor=west] at (0.1,{-0.15-8*0.45-0.30 -0.25 -0.18})
      {\textbf{Sparse Tree Attention Mask}};
  \end{scope}
\end{tikzpicture}}
\caption{\textbf{Efficient verification with sparse tree attention
packing.} The tree of $N$ candidate tokens (left) is packed\revb{, together
with the committed prefix (grey),} into
a verify query of length \revb{$q{=}N{+}1{=}9$} (column headers, top right) and
processed in one \revb{\texttt{MultiCodeDecoder}} model call under the sparse tree
attention mask shown below it (blue: query token $i$ attends to key
position $j$, i.e.\ $j$ is an ancestor of $i$ or $i$ itself).
Row and column headers share the same token ordering. The longest
verified speculation path (green) is committed to the KV cache;
larger $N$ raises both the expected accepted prefix and the verify
cost.}
\label{fig:tree}
\end{figure}

The \texttt{MultiCodeDecoder} is multiheaded by construction and so
requires no architectural changes to support Medusa style speculative decoding,
but it was not \emph{trained} for it: each existing LMH
expects the hidden state at its own codebook position. A direct
application of Medusa style speculative decoding therefore works
but yields low tokens/step. We propose two changes to recover the
gain:


\paragraph{Tree attention increases acceptance rate.}
Instead of drafting a single token per position, the drafter
can emit \revb{the top $m$} candidates at each of the $K$ drafted codebook
positions and arrange them as a tree of depth
$K$\rev{~\cite{cai2024medusa}}:
every node is a draft token, and every root to leaf \emph{path}
is a candidate continuation of length $K$. Taking the Cartesian
product yields up to \revb{$m^{K}$} such paths (the number of candidate
sequences grows exponentially in draft depth), while the tree
itself has only $N$ distinct nodes. These $N$ tokens are packed,
\revb{together with the committed prefix, into a single query of
length $q{=}N{+}1$} and verified in one
\texttt{MultiCodeDecoder} model call, masked so that each token attends
only to its ancestors~\rev{\cite{cai2024medusa,specinfer2024}}
(Figure~\ref{fig:tree}).
\revb{Verification uses the standard speculative acceptance rule
with residual resampling, applied along the
tree~\cite{leviathan2023fast,chen2023accelerating,cai2024medusa}.}
With the original heads this alone raises tokens/step, but the
gain materialises only at large $q$, at which point the verify
model call itself becomes compute bound and erodes the end to end
speedup on older hardware.

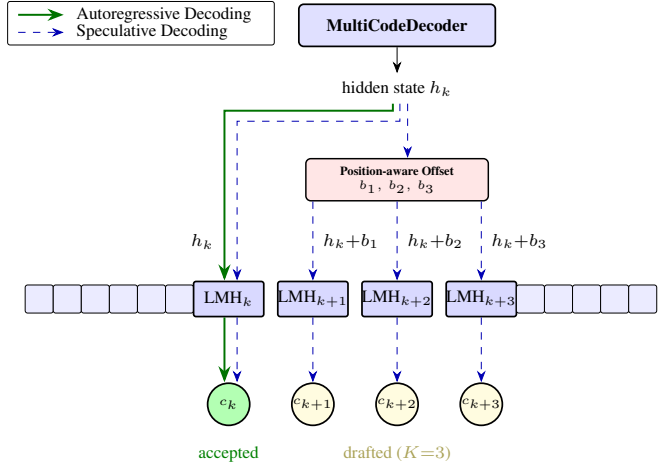
\begin{figure}[h]
\centering
\resizebox{\columnwidth}{!}{%
\begin{tikzpicture}[
    every node/.style={font=\scriptsize},
    head/.style={draw,rounded corners=1pt,minimum width=4mm,
                 minimum height=4mm,inner sep=0pt,fill=blue!8},
    lmh/.style={draw,rounded corners=1pt,minimum width=10mm,
                minimum height=5mm,inner sep=0pt,fill=blue!15,
                line width=0.7pt},
    offsetblk/.style={draw,rounded corners=2pt,inner sep=2pt,
                    fill=red!10,line width=0.6pt,align=center,
                    minimum width=26mm,minimum height=6mm,
                    font=\tiny},
    vftok/.style={draw,circle,minimum size=6mm,inner sep=0pt,
                  fill=green!30,line width=0.7pt,font=\tiny},
    drtok/.style={draw,circle,minimum size=6mm,inner sep=0pt,
                  fill=yellow!15,line width=0.7pt,font=\tiny},
    mcdblk/.style={draw,rounded corners=2pt,inner sep=3pt,
                   fill=blue!12,line width=0.7pt,align=center,
                   minimum width=28mm,minimum height=6mm,
                   font=\scriptsize},
    arrbase/.style={->,>=Stealth,thick,green!45!black},
    arrspec/.style={->,>=Stealth,thin,dashed,blue!70!black}
  ]
  \foreach \xc [count=\i from 0] in {0.0,0.4,0.8,1.2,1.6,2.0} {
    \node[head] (s\i) at (\xc,0) {};
  }
  \node[lmh] (hx)  at (2.7,0) {\scriptsize LMH$_k$};
  \node[lmh] (hx1) at (3.9,0) {\scriptsize LMH$_{k{+}1}$};
  \node[lmh] (hx2) at (5.1,0) {\scriptsize LMH$_{k{+}2}$};
  \node[lmh] (hx3) at (6.3,0) {\scriptsize LMH$_{k{+}3}$};
  \foreach \xc [count=\i from 0] in {7.0,7.4,7.8,8.2,8.6} {
    \node[head] (e\i) at (\xc,0) {};
  }
  \begin{scope}[shift={(-9,4.5)}]
    \draw[draw=black,rounded corners=1pt,fill=white]
      (8.55,-0.88) rectangle (12.35,-0.26);
    \draw[arrbase] (8.70,-0.45) -- (9.35,-0.45);
    \node[font=\scriptsize,anchor=west] at (9.42,-0.45)
      {Autoregressive Decoding};
    \draw[arrspec] (8.70,-0.70) -- (9.35,-0.70);
    \node[font=\scriptsize,anchor=west] at (9.42,-0.70)
      {Speculative Decoding};
  \end{scope}

  \node[mcdblk] (mcd) at (5.1,3.9)
    {\textbf{MultiCodeDecoder}};
  \node[font=\scriptsize] (h) at (5.1,3.0) {hidden state $h_k$};
  \draw[->,>=Stealth,thin] (mcd.south) -- (h.north);

  \node[offsetblk] (off) at (5.1,1.7)
    {\textbf{Position-aware Offset}\\$b_{1},\,b_{2},\,b_{3}$};

  \draw[arrspec] ([xshift=1.5mm]h.south) -- ([xshift=1.5mm]off.north);

  \draw[arrspec] (off.south -| hx1) -- (hx1.north)
    node[midway,right=1pt,font=\scriptsize,text=black!60!black]{$h_k{+}b_1$};
  \draw[arrspec] (off.south -| hx2) -- (hx2.north)
    node[midway,right=1pt,font=\scriptsize,text=black!60!black]{$h_k{+}b_2$};
  \draw[arrspec] (off.south -| hx3) -- (hx3.north)
    node[midway,right=1pt,font=\scriptsize,text=black!60!black]{$h_k{+}b_3$};

  \coordinate (hsL)   at ([xshift=-0.6mm]h.south);
  \coordinate (hsR)   at ([xshift=0.4mm]h.south);
  \coordinate (hxL)   at ([xshift=-0.6mm]hx.north);
  \coordinate (hxR)   at ([xshift=1.2mm]hx.north);
  \coordinate (gDown) at ([yshift=-1mm]hsL);
  \coordinate (gCorn) at (gDown -| hxL);
  \coordinate (bDown) at ([yshift=-2.5mm]hsR);
  \coordinate (bCorn) at (bDown -| hxR);
  \draw[arrbase] (hsL) -- (gDown) -- (gCorn) -- (hxL)
    node[pos=0.77,left=1pt,font=\scriptsize,text=black]{$h_k$};
  \draw[arrspec] (hsR) -- (bDown) -- (bCorn) -- (hxR);

  \node[vftok] (cv) at (2.7,-1.5) {$c_k$};
  \draw[arrbase] ([xshift=-0.6mm]hx.south) -- ([xshift=-0.6mm]cv.north);
  \draw[arrspec] ([xshift=1.2mm]hx.south) -- ([xshift=1.2mm]cv.north);
  \node[font=\scriptsize,green!50!black,below=1.75mm of cv]
    {accepted};

  \node[drtok] (cd1) at (3.9,-1.5) {$c_{k{+}1}$};
  \node[drtok] (cd2) at (5.1,-1.5) {$c_{k{+}2}$};
  \node[drtok] (cd3) at (6.3,-1.5) {$c_{k{+}3}$};
  \draw[arrspec] (hx1.south) -- (cd1.north);
  \draw[arrspec] (hx2.south) -- (cd2.north);
  \draw[arrspec] (hx3.south) -- (cd3.north);
  \node[font=\scriptsize,yellow!60!black] at (5.1,-2.2)
    {drafted ($K{=}3$)};
\end{tikzpicture}}
\caption{\textbf{Comparison of baseline AR and our speculation
inside a single \emph{MultiCodeDecoder} model call (Fig.~\ref{fig:rvqloop}
inner loop).}
All LM heads are original frozen Qwen3-TTS heads. Solid green
arrows denote the baseline autoregressive path: at runtime position
$k$, LMH$_k$ consumes $h_k$ and emits the \emph{accepted} code (green).
Because this first-token path is identical to baseline AR, the code
is guaranteed to be accepted.
Dashed blue arrows denote our speculative path in the same model call:
$h_k$ is fed into a small \textbf{Position-aware Offset} block whose three
outputs $h_k{+}b_1,h_k{+}b_2,h_k{+}b_3$ drive LMH$_{k+1}${--}LMH$_{k+3}$
to emit \emph{drafted} codes (blue) for the next $K{=}3$ RVQ positions
ahead of time.}
\label{fig:offset}
\end{figure}

\paragraph{RVQ position aware offsets with sparse tree attention.}
Because each LMH expects the hidden state at its own codebook
position, projecting a position-$k$ hidden state through
LMH$_{k{+}j}$ does not, on its own, produce strong drafts. A
small per position residual offset on the hidden state recovers the lost
signal (Figure~\ref{fig:offset}): at runtime position $k$,
LMH$_k$ runs without an offset and emits the accepted token, while
LMH$_{k{+}j}$ receives an additively offset hidden state
$h_k{+}b_{j}$, where $b_1,\dots,b_K$ \revb{($K{=}3$)} are trained
per offset residual vectors, and emits one of the $K$ drafts.
\revb{The same $K$ offset vectors are shared across all starting
positions $k$.}
Combined with tree attention drafting, this delivers
most of the tokens/step uplift; to keep the verify model call's
query length small enough to stay bandwidth bound, we further
prune the full tree to a sparse subset of paths (\emph{sparse
tree attention}), retaining most of the acceptance gain while
remaining deployable on older devices.
\revb{The sparse tree is calibrated offline: using per depth and
per rank acceptance statistics measured on the calibration set
(Table~\ref{tab:data_specs}), the tree is grown greedily, at each step
adding the node with the largest expected contribution to the
accepted prefix length~\cite{cai2024medusa}, until the node
budget $N$ is reached; tree depth is capped at $K$. The budget
$N$ is then chosen per device (\S\ref{ssec:device}).}

\rev{\paragraph{Training the offsets.} The backbone, the per position embedding tables, and all LM heads remain frozen; only the residual offset vectors $b_1,\dots,b_K$ (\revb{$3{,}072$} parameters in total; Table~\ref{tab:medusa_ablation}) are trained. We run the frozen model with teacher forcing over the VoxPopuli~\cite{voxpopuli} training split to obtain, at every inner loop position $k$, the hidden state $h_k$ together with the ground truth codes of the subsequent positions. Each offset state $h_k{+}b_j$ is projected through the frozen LMH$_{k{+}j}$ and trained with a cross entropy loss against the ground truth code $c_{k{+}j}$, summed over draft depths $j\le K$ and all valid positions, following the Medusa head training recipe~\cite{cai2024medusa}. \revb{Training uses Adam with a learning rate of $10^{-3}$ and a batch size of $32$ on a single NVIDIA Tesla P100 GPU.} The verifier stream never consumes an offset state, so training the offsets leaves the deployed output distribution untouched. The offsets are trained on English data only; at test time we evaluate on six languages and observe no loss in acceptance (\S\ref{ssec:tps}, Table~\ref{tab:per_lang}), so no multilingual training is required.}


\section{Results}
\label{sec:results}

\subsection{Dataset}
\label{ssec:data}
Table~\ref{tab:data_specs}
summarizes the dataset specifications used across training, calibration, and
evaluation.
\rev{The residual offsets are trained and validated on \revb{disjoint
subsets of the English VoxPopuli train split}. The sparse
tree configuration is then selected on a LibriSpeech \revb{test clean} calibration
set\revb{, following the Medusa tree construction
recipe~\cite{cai2024medusa}: we measure per depth and per rank
acceptance statistics on this set and grow trees greedily under
each node budget $N$ and depth cap $K$, keeping the shape with
the highest expected accepted length (\S\ref{ssec:levers})}.
Finally, acceptance and WER are measured on the Fleurs \revb{train}
split in six languages, of which only English overlaps with the training and
calibration language.}
\revb{All experiments use the $0.6$B CustomVoice variant with the
original sampling configuration~\cite{qwen3tts}; WER is averaged
over ten seeds per system.}

\begin{table}[h]
\centering
\caption{\textbf{Dataset specifications.}
Stage-wise corpora used to train the position-aware residual offsets,
calibrate the sparse tree, and measure acceptance.
English VoxPopuli~\cite{voxpopuli} is split into training and
validation; LibriSpeech~\cite{librispeech} provides the held-out set
on which the sparse tree configuration is selected;
Fleurs~\cite{fleurs} supplies multilingual evaluation across six
languages. \textbf{Frames} count \texttt{CodeDecoder} steps at
$80$\,ms/frame, each frame with 15 \texttt{MultiCodeDecoder} steps.}
\label{tab:data_specs}
\small
\setlength{\tabcolsep}{2pt}
\renewcommand{\arraystretch}{1.05}
\begin{tabular}{@{}lllrrr@{}}
\hline
\textbf{Stage} & \textbf{Dataset} & \textbf{Language} & \textbf{Samples} & \textbf{Hours} & \textbf{Frames}\\
\hline
Training    & VoxPopuli   & English  & 155,513 & $\sim$462 & 20.8M \\
Validation  & VoxPopuli   & English  & 17,200  & $\sim$56  & 2.53M \\
Calibration & LibriSpeech & English  & 1,804   & $\sim$4.4 & 200K  \\
\hline
\multirow{6}{*}{Evaluation}
            & \multirow{6}{*}{Fleurs}
                          & English  & 1,242 & $\sim$3.0 & 135K \\
            &             & German   &   874 & $\sim$2.5 & 112K \\
            &             & Spanish  & 1,216 & $\sim$4.3 & 192K \\
            &             & Russian  &   810 & $\sim$2.3 & 105K \\
            &             & Japanese &   810 & $\sim$2.3 & 102K \\
            &             & Mandarin &   842 & $\sim$2.4 & 106K \\
\hline
\end{tabular}
\end{table}

\subsection{Accuracy: WER parity}
\label{ssec:wer}
Speculative decoding is \emph{distributionally lossless} via the
modified rejection sampling
rule~\cite{leviathan2023fast,chen2023accelerating}, which
preserves the target distribution. \revb{Our residual offsets
enter only the drafter stream (Figure~\ref{fig:offset}), so the
verifier LMH outputs are identical to the AR baseline and the
guarantee applies unchanged. Table~\ref{tab:per_lang} reports the
transcription error of the synthesised audio.}

\rev{The AR and speculative WER columns in Table~\ref{tab:per_lang} are close but not numerically identical: both systems \emph{sample} from the same top-$k$ distribution, so any two runs, speculative or not, differ utterance by utterance, and the reported WER additionally passes through an ASR transcription step (WhisperKit~\cite{whisperkit}) that contributes measurement noise of its own. The speculative WER is no higher than the AR WER in any language; \revb{the differences are small and share a direction, which we leave uncharacterised. Beyond WER, we manually reviewed synthesised audio from both decoders and heard no differences in voice, prosody, or artifacts, and the speculative decoder is deployed in production in \revb{the Argmax SDK~\cite{whisperkit,argmaxsdk}} with no observed quality regressions.}}

\subsection{Theoretical acceleration: tokens/step}
\label{ssec:tps}
\revb{All acceptance rates are measured on free running synthesis,
with the model consuming its own generated codes as in
deployment.}
Table~\ref{tab:medusa_ablation} compares three Medusa style drafter
designs. Reusing the neighbouring LM heads as
drafters costs no parameters but lower acceptance rate. Medusa increases that, at
the cost of \revb{$\sim 94$M new parameters ($+16\%$ on a $0.6$B backbone)}.
Our position aware offsets match Medusa acceptance while adding almost no parameters on top of the original model. Acceptance is also stable across languages
(Table~\ref{tab:per_lang}): although the offsets are trained on
English VoxPopuli~\cite{voxpopuli} and calibrated on English
LibriSpeech~\cite{librispeech}, they deliver
$2.44$--$2.46$ tokens/step on all six
Fleurs~\cite{fleurs} evaluation subsets.

\rev{Table~\ref{tab:medusa_ablation} isolates the effect of the drafter. Naive reuse of the frozen heads (i) yields only $1.70$ tokens/step because of positional mismatch: each frozen LMH$_{k{+}j}$ was trained to read the hidden state at its \emph{own} codebook position, so a position-$k$ hidden state is out of distribution for it. Dedicated Medusa heads (ii) remove this mismatch at the cost of \revb{${\sim}94$M} added parameters. Our offsets (iii) achieve the same effect with a single learned translation of the hidden state per lookahead distance, matching the acceptance of Medusa heads ($2.47$ vs.\ $2.60$ tokens/step) while adding only \revb{$3{,}072$} parameters to the frozen backbone. We attribute the cross lingual stability to the same mechanism: the offsets model the position conditional geometry of the RVQ hierarchy (the codebook to codebook statistics of the frozen audio codec) rather than the lexical content of the English training data, and this structure is largely language agnostic. Finally, Medusa's \revb{${\sim}16\%$} added parameters would also add asset size and memory on device, which the offsets avoid.}

\begin{table}[h]
\centering
\caption{\textbf{Medusa-style drafter ablation.} Three drafter
choices on top of the frozen Qwen3-TTS-0.6B backbone.
(i) \emph{Reused LM heads}: the $K{=}3$ neighbouring codebook heads
$\text{LMH}_{k{+}1},\dots,\text{LMH}_{k{+}3}$ draft from the
position-$k$ hidden state; no new parameters.
(ii) \emph{Medusa heads}~\cite{cai2024medusa}: $K{=}3$ fresh Medusa
heads per codebook position (\revb{$45$} new heads total).
(iii) \emph{Position-aware offsets} (ours): \revb{$K{=}3$} residual bias
vectors $b_1,\dots,b_K$ added to the hidden state; LM heads frozen
and reused as drafters.}
\label{tab:medusa_ablation}
\small
\setlength{\tabcolsep}{5pt}
\renewcommand{\arraystretch}{1.05}
\begin{tabular}{@{}lcc@{}}
\hline
\textbf{Drafter} & \textbf{Toks/step} & \textbf{Added params} \\
\hline
Reused LM heads                  & 1.70 & --- \\
Medusa heads~\cite{cai2024medusa} & 2.60 & \revb{$\sim 94$M ($\sim {+}16\%$)} \\
\textbf{Ours} (pos.\ offsets)    & \revb{\textbf{2.47}} & \revb{$\mathbf{3{,}072}$ ($\sim{+}5{\times}10^{-4}\%$)} \\
\hline
\end{tabular}
\end{table}


\begin{table}[H]
\centering
\caption{\textbf{Per-language performance on Fleurs.} Acceptance
rate (\emph{Toks/step}) and WER \revb{(CER for Mandarin and
Japanese), in percent,} of the generated audio for the
autoregressive baseline and our speculative decoder, evaluated on
Fleurs subsets in six languages. WER is computed via
WhisperKit~\cite{whisperkit}, using
whisper-large-v3-turbo~\cite{whisper} for Mandarin and Japanese and
parakeet-v3~\cite{parakeet} otherwise.
\revb{Transcription and WER evaluation use our open source
OpenBench suite~\cite{openbench,sdbench}.} As shown in prior
work~\cite{leviathan2023fast,chen2023accelerating}, speculative
decoding is distributionally lossless; \revb{the WER columns are
consistent with this.}}
\label{tab:per_lang}
\small
\setlength{\tabcolsep}{2.4pt}
\renewcommand{\arraystretch}{1.05}
\begin{tabular}{@{}lccccc@{}}
\hline
\multirow{2}{*}{\textbf{Language}}
  & \multirow{2}{*}{\textbf{\shortstack{Audio\\(min)}}}
  & \multicolumn{2}{c}{\textbf{Autoregressive}}
  & \multicolumn{2}{c}{\textbf{Speculative}} \\
\cline{3-4}\cline{5-6}
  &
  & \textbf{Toks/step} & \revb{\textbf{WER/CER}}
  & \textbf{Toks/step} & \revb{\textbf{WER/CER}} \\
\hline
English  & 180.2  & \textit{1.00} & 4.25 & 2.45 & 3.99 \\
German   & 149.0  & \textit{1.00} & 6.4  & 2.46 & 5.58 \\
Spanish  & 255.6  & \textit{1.00} & 7.28 & 2.44 & 5.98  \\
Russian  & 139.8  & \textit{1.00} & 9.55  & 2.46 & 8.26 \\
Japanese & 136.2  & \textit{1.00} & 8.55  & 2.46 & 7.71  \\
Mandarin & 141.7  & \textit{1.00} & 7.43  & 2.45 & 7.18  \\
\hline
\end{tabular}
\end{table}

\subsection{Practical acceleration: per device wall time latency}
\label{ssec:device}
\begin{figure}[h]
\centering
\begin{tikzpicture}
\begin{axis}[
    width=\columnwidth,
    height=0.72\columnwidth,
    xlabel={Sparse-tree Size (N)},
    ylabel={Speed-up ($\times$)},
    xmin=10, xmax=70,
    ymin=1.0, ymax=3.0,
    xtick={16,32,48,64},
    ytick={1.0,1.5,2.0,2.5,3.0},
    legend cell align=left,
    legend style={font=\fontsize{7}{8.5}\selectfont,draw=none,
                  fill=none,
                  inner sep=1pt,row sep=-4pt,
                  at={(0.02,0.98)},anchor=north west},
    legend image post style={scale=0.65},
    grid=major,
    grid style={dotted,gray!40},
    tick label style={font=\footnotesize},
    label style={font=\footnotesize},
]
\addplot[forget plot,color=black!55,thick,dashed,mark=none]
    coordinates {(16,2.25) (32,2.47) (48,2.55) (64,2.60)};

\addplot[forget plot,color=blue!65!black,thick,mark=*,
         mark options={scale=0.85}]
    coordinates {(16,2.21) (32,2.34) (48,2.07) (64,2.05)};

\addplot[forget plot,color=purple!70!black,thick,mark=square*,
         mark options={scale=0.75}]
    coordinates {(16,2.06) (32,2.13) (48,1.80) (64,1.82)};

\addplot[forget plot,color=orange!95!black,thick,mark=triangle*,
         mark options={scale=1.0}]
    coordinates {(16,1.82) (32,1.48) (48,1.32) (64,1.21)};



\addlegendimage{color=blue!65!black,thick,mark=*,
                mark options={scale=0.85}}
\addlegendentry{iPhone 17 Pro Max}

\addlegendimage{color=purple!70!black,thick,mark=square*,
                mark options={scale=0.75}}
\addlegendentry{MacBook M3 Pro}

\addlegendimage{color=orange!95!black,thick,mark=triangle*,
                mark options={scale=1.0}}
\addlegendentry{iPhone 12}

\addlegendimage{color=black!55,thick,dashed,mark=none}
\addlegendentry{Theoretical (Hardware-agnostic)}
\end{axis}
\end{tikzpicture}
\caption{\textbf{Practical speedup vs sparse tree size across
hardware.} The grey dashed curve is the device independent acceptance
(\emph{Toks/step}) and is the theoretical ceiling on the speedup.
Coloured curves are the per device \texttt{MultiCodeDecoder} speedup,
$(\text{Toks/step})/(1+\text{verify overhead})$; the gap below the
ceiling is the verify-step overhead introduced when the query length
grows from $q{=}1$ (AR, memory bound) to $q{=}N$ (compute bound).
\rev{The selected operating point is $N{=}32$ on iPhone~17 Pro Max
and MacBook M3~Pro, and $N{=}16$ on iPhone~12.}}
\label{fig:tps_device}
\end{figure}
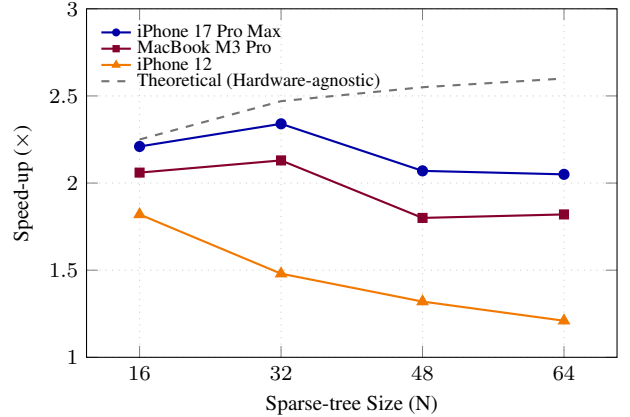

Acceptance rate (\emph{Toks/step}) is the ceiling for the practical speedup; the
realised speedup is always lower because the verify call runs at
query length $q{=}N$ rather than $q{=}1$; as $N$ grows this pushes
the \texttt{MultiCodeDecoder} from memory bandwidth bound toward
compute bound. Figure~\ref{fig:tps_device} sweeps $N$
across three Apple devices and exposes the per device optimum.

\rev{Figure~\ref{fig:tps_device} quantifies this tradeoff
(all latencies are p50 over $200$ steady state iterations on the
Apple Neural Engine, W8A16, with $20$ warmup calls excluded).
\revb{W8A16 (8 bit weights, 16 bit
activations)~\cite{whisperkit} is used for all reported
results.} The
acceptance ceiling (grey dashed) rises monotonically with
sparse tree size but saturates (from $2.25$ tokens/step at
$N{=}16$ to $2.60$ at $N{=}64$), whereas the verify cost grows
steadily with the query length $q{=}N$, so the net speedup peaks
where the marginal accepted token no longer pays for the marginal
verify cost. On recent hardware the verify call stays close to memory
bandwidth bound through $N{=}32$: one AR
\texttt{MultiCodeDecoder} call costs $2.05$\,ms on iPhone~17 Pro
Max and $2.2$\,ms on MacBook M3~Pro, and the speculative
call at $N{=}32$ adds only $+5.3\%$ and $+13.6\%$
respectively, so the optimum sits at $N{=}32$ with
$2.34\times$ and $2.13\times$ speedup. On iPhone~12 the crossover
into the compute bound regime occurs almost immediately: the
optimum is $N{=}16$ ($1.82\times$), and larger trees erode the
gain, degrading to $1.21\times$ at $N{=}64$. $N$ can therefore be
calibrated once per device class, offline: acceptance (the
numerator) is device independent, so only the denominator, the
verify overhead, needs profiling on new hardware.}

\revb{The overhead in Figure~\ref{fig:tps_device} isolates the
model call ($+5.3\%$ and $+13.6\%$ at $N{=}32$); host side work
outside the model call brings the total per round overhead to an
estimated $10$--$20\%$, consistent with the $2$--$2.2\times$
deployed speedup range.}

\rev{The absolute latencies show that speculation is most
beneficial on older devices. At $187.5$ inner loop calls per second of audio,
autoregressive token generation alone costs ${\sim}894$\,ms of
inner loop compute per second of synthesised speech
on iPhone~12 ($4.77$\,ms per call), nearly the entire
real time budget before the $28$-layer \texttt{CodeDecoder} and
the speech decoder are accounted for. Speculation at the
selected operating point roughly halves this cost, leaving
headroom for the unaccelerated components. On recent
devices the same reduction cuts this cost from
${\sim}384$\,ms per second of audio (iPhone~17 Pro Max) to
${\sim}164$\,ms, after which token generation accounts for a
minority of the single stream compute budget.}


\section{Conclusion}
\label{sec:conclusion}
\rev{We presented RVQ position aware speculative decoding for
on device TTS, which reuses the frozen per position LM heads of the
Qwen3-TTS \texttt{MultiCodeDecoder} as drafters: \revb{$3{,}072$} trained
residual offset parameters let each head draft ahead of its own
codebook position, while sparse tree attention fuses drafting and
verification into a single compiled model call that leaves the
verifier stream untouched and remains distributionally lossless
under \revb{top-$k$} sampling. The method sustains $2.44$--$2.46$ accepted
tokens per model call across the six Fleurs
evaluation languages with WER parity against the autoregressive
baseline, and delivers $2$--$2.2\times$ RVQ token generation
speedup \revb{on recent} iPhone
and Apple Silicon Mac devices in the single stream regime. On
recent devices this adds headroom; on older hardware such as
iPhone 12, where autoregressive decoding alone nearly
exhausts the real time budget, it unlocks comfortable real time
synthesis. The unaccelerated
$28$-layer
\texttt{CodeDecoder} is a natural next target for speculation across
frames, alongside dynamic per device selection of the sparse tree
size $N$. \revb{The method assumes only an autoregressive codebook depth
decoder with reusable per position heads, a structure shared by
systems such as CaT-TTS~\cite{cattts}; evaluating such transfer
is left for future work.}}

\section{Acknowledgments}
\label{sec:ack}
\revb{This work was funded by Argmax, Inc. All authors are
employees of Argmax, Inc. The authors have no other
relevant financial or non-financial interests to disclose.
Anthropic's Claude~\cite{anthropic_claude} was used to draft and
edit the manuscript text, and OpenAI's
GPT-6~\cite{openai_gpt6} was used to review and critique earlier
drafts. All content was reviewed and verified by the authors, who
take full responsibility for it.}

\section{Compliance with Ethical Standards}
\label{sec:ethics}
\revb{This study uses only the publicly released
VoxPopuli~\cite{voxpopuli}, LibriSpeech~\cite{librispeech}, and
Fleurs~\cite{fleurs} datasets under their respective licenses and
collects no new human subjects data, so no ethical approval was
required.}

\bibliographystyle{IEEEbib}
\bibliography{main}

\end{document}